\documentclass{article}

\usepackage{arxiv}
\usepackage[utf8]{inputenc} 
\usepackage[T1]{fontenc}    
\usepackage{hyperref}       
\usepackage{url}            
\usepackage{booktabs}       
\usepackage{amsfonts}       
\usepackage{nicefrac}       
\usepackage{microtype}      
\usepackage{lipsum}		
\usepackage{graphicx}
\usepackage[numbers]{natbib}
\usepackage{doi}
\usepackage{caption} 
\title{Brexit and bots: characterizing the behaviour of automated accounts on Twitter\\ during the UK election}

\author{
Matteo Bruno$^*$\\
	IMT Lucca\\
	Piazza S. Francesco 19, 55100\\
	Lucca, Italy\\
	$^*$\texttt{matteo.bruno@imtlucca.it} \\
	\And
	Renaud Lambiotte\\
	Mathematical Institute\\
	University of Oxford\\
	Woodstock Road\\
	Oxford OX2 6GG, UK\\
	\And
	Fabio Saracco\\
	Institute for Applied Mathematics\\
	National Research Council\\
	Via dei Taurini 19, 00185\\
	Rome, Italy\\
	And\\
	IMT Lucca\\
	Piazza S. Francesco 19, 55100\\
	Lucca, Italy\\
}

\date{}

\renewcommand{\headeright}{}
\renewcommand{\undertitle}{}

\hypersetup{
pdftitle={Brexit and bots: characterizing the behaviour of automated accounts on Twitter during the UK election},
}

\begin{document}
\maketitle

\begin{abstract}
Online Social Networks represent a novel opportunity for political campaigns, revolutionising the paradigm of political communication. Nevertheless, many studies uncovered the presence of d/misinformation campaigns or of malicious activities by genuine or automated users, putting at severe risk the credibility of online platforms. This phenomenon is particularly evident during crucial political events, as political elections. In the present paper, we provide a comprehensive description of the structure of the networks of interactions among users and bots during the UK elections of 2019. In particular, we focus on the polarised discussion about Brexit on Twitter analysing a data set made of more than 10 million tweets posted for over a month. We found that the presence of automated accounts fostered the debate particularly in the days before the UK national elections, in which we find a steep increase of bots in the discussion; in the days after the election day, their incidence returned to values similar to the ones observed few weeks before the elections. On the other hand, we found that the number of suspended users (i.e. accounts that were  removed by the platform for some violation of the Twitter policy) remained constant until the election day, after which it reached significantly higher values. Remarkably, after the TV debate between Boris Johnson and Jeremy Corbyn, we observed the injection of a large number of novel bots whose behaviour is markedly different from that of pre-existing ones. Finally, we explored the bots' stance, finding that their activity is spread across the whole political spectrum, although in different proportions, and we studied the different usage of hashtags by automated accounts and suspended users, thus targeting the formation of common narratives in different sides of the debate.
\end{abstract}


\section{Introduction}\label{intro}
Social networks are becoming more and more important in the modern world, and research has accordingly focused on the description and analysis of their structure and dynamics~\cite{Adamic2005,bessi2016social,DelVicario2016,ferrara2016rise,Lazer2018}. 
While the effective impact on election results is still debated, as it is on public opinion about pivotal political topics~\cite{Dubois2018}, it is nevertheless undeniable that new media represent an important tool to shape the political communication strategy.
For this reason, it is of crucial importance to investigate possible artificial manipulations of data and of the behaviour of users. Malicious behaviours have been identified in recent years, such as the employment of automated accounts to foster effects like echo chambers \cite{stella2018bots}, push some argument towards one direction or increase the visibility and credibility of users \cite{broniatowski2018weaponized,bessi2016social,ferrara2016rise,caldarelli2020role}. The injection of coordinated accounts is particularly common during elections and key events in several countries \cite{shao2017spread,pastor2020spotting,rossi2020detecting}. Detecting the bots and analysing the behaviour of malicious accounts is not, however, an easy task. For now more than a decade, scientists have worked on the detection of automated accounts, using  a variety of approaches involving machine learning to analyse language and behavioural features \cite{lee2010uncovering,ferrara2016rise,cresci2017social} and the analysis of social interactions \cite{mehrotra2016detection,jia2017random}.

In this paper, we focus on the discussion about Brexit on Twitter before and during the UK elections of 2019. The presence and influence of automated accounts in the Twitter Brexit discussion was already noted  during the 2016 referendum. 
In their work, Bastos and Mercea \cite{bastos2017} found a botnet of more than 10k bots, propagating fake and hyperpartisan news. They found that the bots are specialising in different activities, with some of them retweeting active users and some other generating cascades of information from other bots' tweets. Similarly, the work of Howard and Kollanyi \cite{howard2016bots} showed that the bots were mainly focused on retweeting and that their activity was very high compared to genuine users. Furthermore, they found that the bots' activity was more in support of Brexit than against it.
Here, we focus on the UK elections in 2019, as they played a  pivotal role for the confirmation of Brexit and as they polarised the public opinion about the future relations of the UK with the European Union.
To do so, we have analysed a dataset that includes more than 10 million Brexit-related tweets and 1 million users, over a period of more than one month centred on the day of the general election, held on the 12th of December 2019.

Our main contribution is the  quantification of the bots' presence in the discussion and the characterisation of their behaviour. We uncover a massive participation of bots, and we discover a large addition of bots in the discussion right before the general elections. Moreover, we show evidence that the bots that are seemingly injected in the discussion behave in a very different manner to those that were already present. We analyse the accounts that have been suspended as well, finding non-trivial temporal patterns of their activity and confirming some of the results presented in Chowdhury et al. \cite{chowdhury2020twitter}.
In addition, we also identify specific patterns of retweets by bots and suspended users, and we are able to find network metrics that differentiate automated accounts and humans. Finally, we uncover the topics in which the bots were active by analysing their retweet activity. We quantify the presence of bots and suspended users in different communities of the retweet network. Using network methods, we were also able to find significant groups of bots specialised in retweeting similar kinds of hashtags and URLs.

\section{Methods}\label{sec:methods}

\subsection{Data collection}
We downloaded our data using Tweepy\footnote{\url{https://www.tweepy.org/}}, a Python wrapper for the official Twitter API. The period that we considered goes from the 20th of November 2019 to the 23rd of December 2019 (the UK elections were held on the 12th of December). During this period, we  registered all the tweets that contained the keyword \emph{Brexit}. 
It could be argued that our filtering procedure is not adequate to analyze the whole discussion, as the UK general election also covered other topics of discussion that Brexit. However, our aim was  to investigate the underlying sub-discussion on Brexit, as it played a central and polarising role, but also gained global attention from all over the world. 

\subsection{Automated and suspended accounts}
To assign a bot score to each account, we used the Botometer API \cite{davis2016botornot} in its most recent v4 version \cite{Sayyadiharikandeh2020}. 
Where not otherwise specified, all unverified users\footnote{Twitter has a procedure to check the identity of some users that are of public interests, as the official account of political parties, newspaper, online newscasts, politicians, journalists and  VIPs in general. Due to the verification procedure of Twitter, all verified accounts are considered as genuine by Botometer.} with a Botometer CAP (Complete Automation Probability) score greater or equal than 0.43 are considered bots, as was already chosen in Ref. \cite{varol2017online}. The users that had been suspended and removed from Twitter after a short time, thus being not classifiable, have been treated separately. There is no strict consensus on the threshold value that should be chosen, and it is sometimes taken higher; however our data shows that setting a higher value is not appropriate given our short time window. In our case, we chose the threshold so that the top 7 percentile of users will be labeled as bots. A similar approach was adopted by Ferrara et al. \cite{ferrara2020covid}. In Appendix~\ref{app:cap_discussion} we show that the choice of the threshold does not modify the observed behaviour of the group, even if it changes the total number of automated accounts.

Furthermore, we decided to treat suspended users differently. In our analysis, we encountered a high number of accounts that were suspended (the Twitter API does not give more information). Dealing with these users, it is not possible to examine their past activity and give them a bot score, so we only had the posts that we downloaded. We decided to treat them as a separate class, because it is not granted that a suspended user was suspended for violating the policies of Twitter on automation. The analysis presented in \cite{chowdhury2020twitter} showed that the activity of suspended users presents many traits of malicious behaviours.

\subsection{Network models}\label{sec:methods_network_models}

\emph{Network projection.} In order to analyse the bipartite networks we built with our data, we apply a recently-proposed algorithm to obtain monopartite representations of bipartite networks \cite{Saracco2016b}. The procedure consists in calculating, for each couple of nodes in one layer, the number of common neighbors on the opposite layer as a measure of similarity. Next, the observed similarities are compared to the expectations of an entropy-based null model that accounts for the nodes' degrees: if the observed similarity between the two nodes is statistically significant, a link connecting the two nodes is present in the one-mode projection.\\
We use the implementation proposed in \cite{vallarano2021} and the relative code of the package NEMtropy\footnote{\url{https://github.com/nicoloval/NEMtropy}} for the computation of the randomization and of the monopartite projection. In the following, the nodes corresponding to one layer will be called with the letter $r=1,\dots ,R$ as they correspond to the rows layer in a biadjacency matrix representation of the network, and the opposite layer will be indexed as $c=1,\dots ,C$ as they correspond to columns. We will call the biadjacency matrix representing the network with the letter $\mathbf{B}$ and its entries with $b_{rc}$. 

In a simple one-mode projection, we can produce a monopartite network of the nodes of one layer of a bipartite network by calculating the number of neighbors they share and setting that measurement as the weight of the link between them. The number of common neighbors is: 

\begin{equation}\label{eq:V}
V_{rr'}=\sum_{c}b_{rc}b_{r'c}=\sum_{c}V_{rr'}^c.
\end{equation}
where we use $V_{rr'}^c\equiv b_{rc}b_{r'c}$ meaning that $V_{rr'}^c=1$ if and only if $c$ shares a link to both $r$ and $r'$. In our approach, we compute this as a measure of the similarity of the nodes.

We then want to compare the measure of the similarity of the nodes to the similarity that is expected from a null model that accounts for the nodes' degrees. As a benchmark, we use the Bipartite Configuration Model (BiCM \cite{Saracco2015}), that belongs to the family of entropy-based null-models, \cite{Park2004,Garlaschelli2008,Squartini2011,Fronczak2012}. These models arise from the maximisation of the Shannon entropy, which corresponds to maximising the uncertainty about the system, given some constraints, and resulting in a maximally unbiased model. Practically, the constrained maximisation of the Shannon entropy $S=-\sum_{\mathbf{B}}P(\mathbf{B})\ln P(\mathbf{B})$ leads to an exponential form for the probability distribution 
\begin{equation}
P(\mathbf{B})=\frac{e^{-H(\vec{\theta},\:\vec{C}(\mathbf{B}))}}{Z(\vec{\theta})},
\end{equation}
 where $\mathbf{C}(\mathbf{B})$ is the vector of constraints evaluated on the biadjacency matrix $\mathbf{B}$, $\vec{\theta}$ is the vector of the Lagrangian multipliers associated to the maximisation procedure, $Z(\vec{\theta})$ is the partition function and $H(\vec{\theta},\:\vec{C}(\mathbf{B}))$ is the Hamiltonian associated to the maximisation problem~\cite{Park2004}.
 For each different case, the actual numerical probability $P(\mathbf{B})$ is determined by the targeted topological constraints that need to be accounted for \cite{Park2004}.
In our case, we need to verify that the similarity between two nodes will be statistically significant when considering the respective degrees, so we employ the BiCM, that is the bipartite entropy-based null-model that takes as constraints the degree sequences of the two layers. With these constraints, the probability of a graph factorises in terms of probabilities per link:

\begin{equation}
P(\mathbf{B})=\prod_{r=1}^R\prod_{c=1}^Cp_{rc}^{m_{rc}}(1-p_{rc})^{1-m_{rc}}.
\end{equation}
Each link probability will be a function of the $(R+C)$-vector of Lagrange multipliers $\vec{\theta}$ associated to the degrees as 
\begin{equation}
p_{rc}=\frac{x_ry_c}{1+x_ry_c},
\label{prob}
\end{equation}
where $x_r=e^{-\theta_r}$ and $y_c=e^{-\theta_c}$ are reparameterizations of the Lagrange multipliers.

In order to find the numerical value of the probability, we  set the average degrees of the model to the observed ones and solve the set of nonlinear equations

\begin{equation}\label{sys}
\left\{ 
\begin{array}{ll}
\langle k_r\rangle=\sum_c p_{rc} = k_r^*,\:\: r=1\dots R\\
&\\
\langle h_c\rangle=\sum_r p_{rc} = h_c^*,\:\: c=1\dots C
\end{array}
\right.,
\end{equation}
where $k_r$ and $h_c$ are the degree of the node $r$ and $c$ respectively, and $^*$ indicates the observed empirical value. More details can be found in \cite{vallarano2021}.

The entropy-based null-model model allows to treat links as independent random variables, so the probability of a common neighbor reads $P(V_{rr'}^c=1)=p_{rc}p_{r'c}$. The distribution of a $V_{rr'}$ is then a sum of independent Bernoulli random variables with different coefficients, resulting in a Poisson-Binomial distribution. We then measure the statistical significance of the similarity of two nodes $r$ and $r'$ by calculating the p-value of the observed $V_{rr'}^*$ with respect to the so obtained distribution. To perform this computation, given the large size of our datasets, we approximate the Poisson-Binomial distribution with a Poisson one with the same mean. The error of this approximation is controlled by Le Cam's theorem \cite{Deheuvels1989,Volkova1996,Hong2013}:

\begin{equation}
\sum \limits_{k=0}^C\left\vert f_{PB}(V_{rr'}=k)-\frac{\mu^k \exp(-\mu)}{k!}\right\vert<2\sum_{c=1}^C(p_{rc}p_{r'c})^2.
\label{Poisson approximation}
\end{equation}

To validate links, we set a threshold that will distinguish the p-values that are significant from the ones to be disregarded. We apply the False Discovery Rate (FDR) procedure \cite{Benjamini1995}. With $H_1, \dots H_N$ different hypotheses associated to different p-values, the recipe of FDR prescribes to sort the p-values in increasing order and than look for the largest integer $i$ such that 
\begin{equation}
\textmd{p-value}_{\hat{i}}\leq\frac{\hat{i}t}{N}
\label{threshold}
\end{equation}
where $t$ is a single-test significance level (we use $t=0.05$ unless otherwise stated). Than we reject all hypotheses $H_i$ such that $i<\hat{i}$.
In our case, the procedure validates all links with a similarity p-value higher than the threshold found via the FDR.

\begin{figure}[ht!]
\centering
\resizebox{\textwidth}{!}{%
 \includegraphics{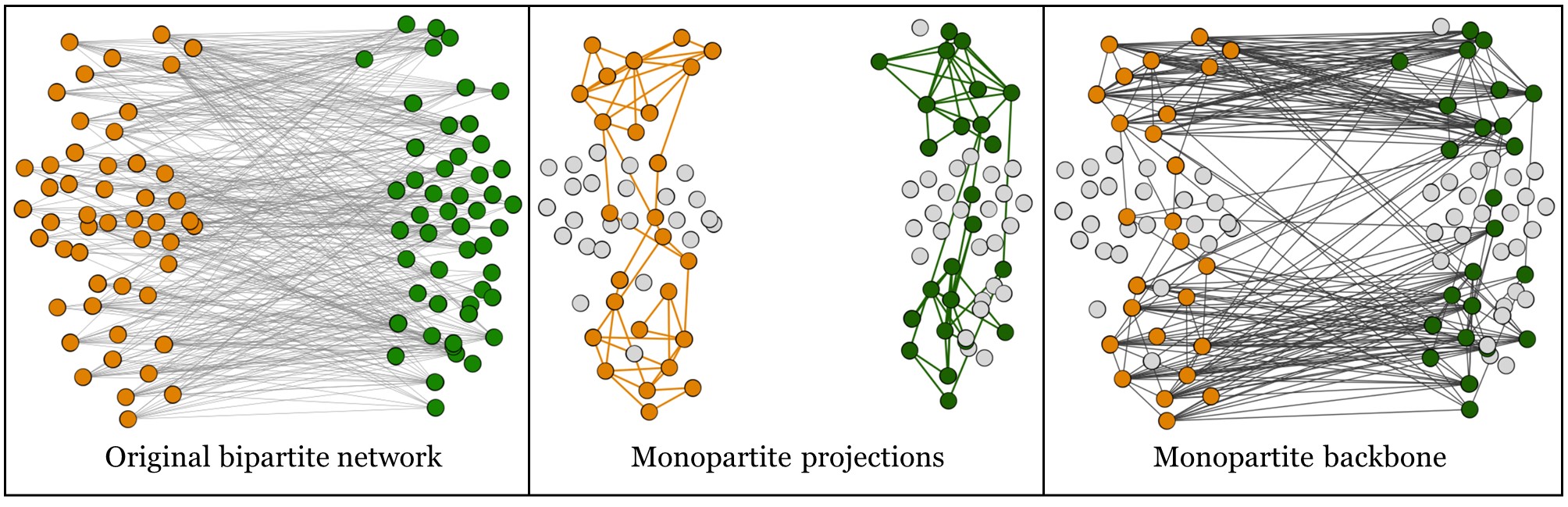}
}
\caption{\textbf{Monopartite backbone extraction from a bipartite network}. The bipartite network of the first panel is projected on both layers. After the two one-mode projections are obtained (second panel), the original links can be added again to obtain the backbone of the network (third panel), that will highlight mixed groups of interactions.}
\label{fig:backbone_extraction}
\end{figure}

\emph{Monopartite backbone.} After obtaining the two one-mode projections of a bipartite network, we add again the links of the original bipartite network to connect the nodes that remained in the projections. This way, we obtain a monopartite network that will show the groups of similar nodes of the same type, and link the groups if they were originally related. A representation of this procedure is shown in Fig. \ref{fig:backbone_extraction}.

\emph{Discursive communities.} In~\cite{becatti2019extracting} a method for extracting memberships of users to different discursive communities in social networks was presented and later refined in \cite{caldarelli2020role,Caldarelli2021}. The method is based on the behaviour of verified users, i.e. the accounts that are certified by a Twitter official procedure: these are property of newspapers, newscasts, journalists, politicians, political parties and VIPs in general that can be of public interest. Verified users have a stronger tweeting (i.e. generation of original contents) than retweeting (i.e. sharing posts published by others) activity, on average. In this sense, we can leverage on the presence of verified users to extract discursive communities. In fact, we can use the validated projection described earlier and apply it to a bipartite network of verified and unverified users, in which a link is present if one of the users retweeted the other one at least once. In this sense, we are stating that two verified users are perceived as similar if they both interact with a considerable number of unverified accounts. After obtaining the validated projection, we can find different communities of verified users using the Louvain algorithm. With the so found community labels, we can obtain groups of similar unverified users in the original network by applying a label propagation algorithm. In this work, we simply label a user with the most common label among its neighbours, removing one edge at random in case of ties, using the algorithm proposed in \cite{raghavan2007near} and considering the labels of verified users as fixed. Iterating the procedure several times until consensus is reached and no node will change its label, we obtain a labeling for all users. Since this procedure can yield different results for the random choice we make when breaking ties, we repeat the whole label propagation 1000 times and take the most likely outcome. More details on this procedure can be found in~\cite{becatti2019extracting,caldarelli2020role,Caldarelli2021}.

\subsection{Community detection}
To investigate the community structure of the networks, we use modularity-optimisation strategies. For monopartite networks, we look for optimal modularity partitions using the Louvain algorithm \cite{Blondel2008}. The Louvain community detection algorithm is order dependent~\cite{Fortunato2010}, so we rerun the algorithm after reshuffling the order of the nodes. We then consider the partition in community displaying the largest value of the modularity.

\subsection{Core-periphery measures}
To measure if nodes belong to the core or periphery of their communities, we employ a strategy similar to the one found in Guimer\'a et al. \cite{guimera2005}. We use two measures: the first one, coming directly from this original paper, is called \emph{participation score} and reads
\begin{equation}\label{participation_score}
    P(i) = 1 - \sum\limits_{c=1}^{C}\left(\frac{k_{ic}}{k_i}\right)^2
\end{equation}
where $i$ is a node of the graph, $C$ is the number of communities, $k_i$ is the degree of node $i$ and $k_{ic}$ is the degree of node $i$ towards the community $c$, i.e. the number of neighbors of $i$ belonging to community $c$. The participation score will be $0$ for a node that has all its neighbors in its own community, and $\sim \frac{c_i^2 - 1}{c_i^2}$ if its neighbors are all equally distributed inside $c_i$ communities.

We also evaluate how relevant a node is inside its community, in accordance to the strategy of \cite{guimera2005} but using a slightly different measure. In order to compare the degree of the nodes towards other nodes of their community, and given the non-gaussian distribution of the in-degrees inside a community, we adopt the following \emph{relevance score}:
\begin{equation}\label{relevance_score}
    R(i) = - \log \textmd{pval} (d_{ic_i}) = - \log \mathbf{P}(d_{jc_i}\ge d_{ic_i})
\end{equation}

\section{Results}\label{sec:results}

\begin{figure}[htb!]
\centering
\resizebox{\textwidth}{!}{%
 \includegraphics{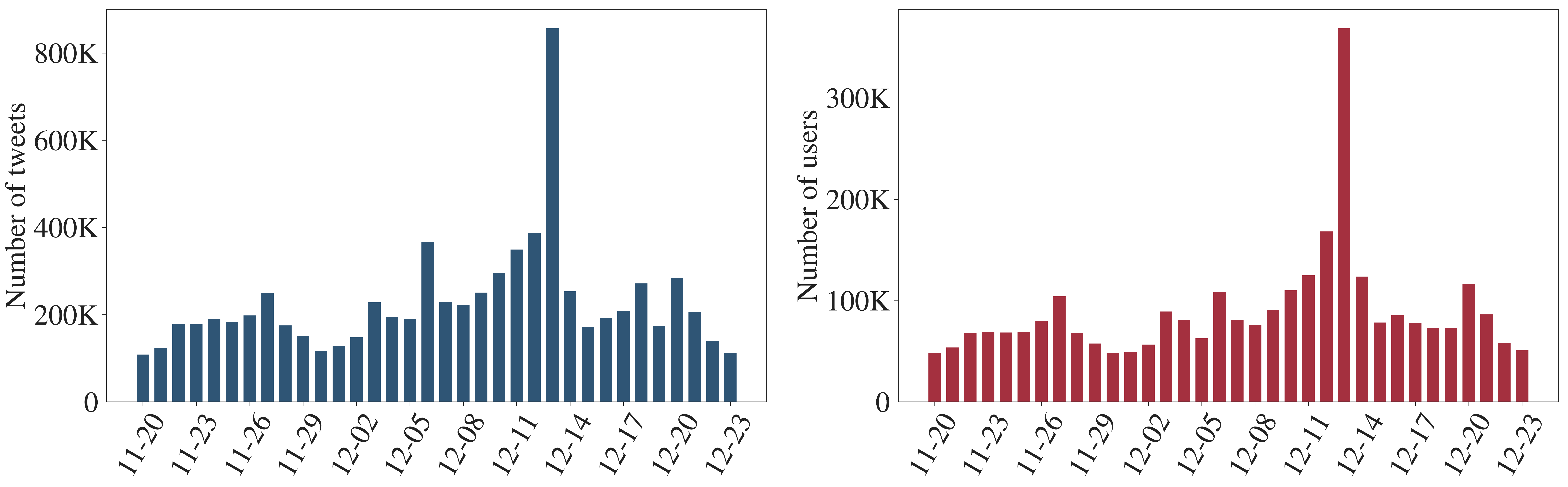}
}
\caption{\textbf{Number of tweets (including retweets) and users per day.} The peak on the 12th of December that can be observed in both panels is in concurrence of the day of the elections.}
\label{fig:tweets_num}
\end{figure}

The number of posts per day fluctuates between 100K and 300K for most of the time period (Fig. \ref{fig:tweets_num}), and it jumps to 900K on  election day. In total, we analysed more than 10 million posts. The number of users follows a similar trend, being almost never lower than 50K and rising as high as 350K on the the election day, for a total number of over one million users.

\subsection{Bots statistics}

\begin{figure}
\centering
\resizebox{\textwidth}{!}{%
 \includegraphics{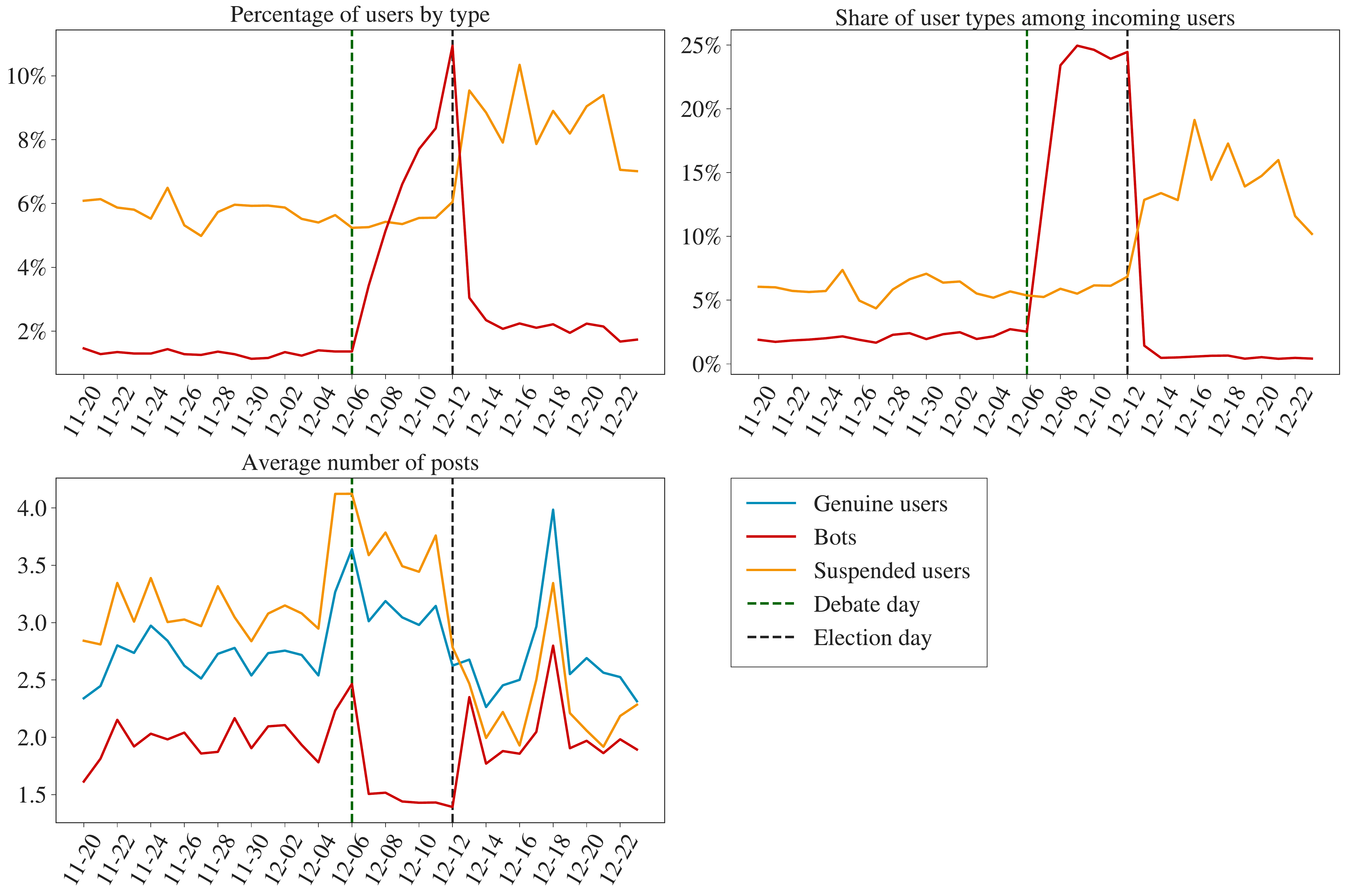}
}
\caption{\textbf{Presence of bots and suspended users in the discussion over time.} For users labeled as bots (with a score higher than 0.43) that have not been suspended, there is a change after the debate of the 6th of November (top left), with new bots coming into the discussion (top right). The new bots seem less active than the old ones (bottom). The percentage of bots among genuine users becomes as high as 10\%. Among the removed users, the changes happen after the 12th of December (election day), with new suspended users entering the discussion while being less active on average.}
\label{fig:bots_percentage}
\end{figure}

The average percentage of bots in the discussion fluctuates around 2 percent for the first part of our time period (Fig. \ref{fig:bots_percentage}). Interestingly, though, the number of bots in the dataset increases steeply after the 6th of December. It rises as high as 11 percent on the election day, almost 10 percent higher than one week before, but rapidly drops after the election. The number of suspended users also shows a change of behaviour around that date, but a different one: their percentage goes up only moderately, from 6 percent to 9 percent after the election day, and this increase is not followed by a rapid drop. It can be observed that the increase in the bots' presence comes with a relative decrease in their activity, as per Fig. \ref{fig:bots_percentage}, third panel: they were on average tweeting more before the 6th of December, while the separation between users and bots increases from the 8th and they start to tweet less in the period of their increase before the elections. This behaviour comes mostly from retweet activity, so the bots that enter the Brexit discussion after the 6th of December seem to be more silent than the ones that were already present in the dataset.

For suspended users, the activity trends are almost inverted: they are more active than a normal user and way more than a bot, perhaps making it easier for Twitter to detect such accounts. Their activity does not show the same pattern of the bots but instead goes down after the election day.

As shown in Table \ref{tab:user_type_retweets}, bots and suspended users tend to retweet malicious accounts more than genuine users, but they are not very supportive of each other and their activity focuses mainly on retweeting genuine accounts, thus generating noise and fostering discussions that are already present. Also, they are not much retweeted or quoted by genuine users, with only around 4\% of the retweet/quote activity of genuine unverified users interacting with suspended users or bots, and even less for verified accounts. This suggests that the bots' strategy is to go under the radar by not generating large cascades of false information, but rather limiting their activity to increase the credibility and visibility of genuine accounts.

The quantity of new bots that appear in the discussion can be seen in the second panel of Fig. \ref{fig:bots_percentage}. The figure focuses on the new users and bots that appear for the first time in our dataset. We have some data from the previous days so the plots are rather stable.
The increase in the percentage of bots and suspended users comes with a higher relative percentage of new users of the corresponding type introduced in the discussion compared to the quantity of new users. This rapid, and large, increase of a new kind of bots in the dataset appears to be rather unique. With our analysis, we found differences between the Twitter activity of the already present bots and of the new bots. We also found significant differences between bots and suspended users.

\begin{table}[ht!]
    \centering
    \begin{tabular}{|c | c | cccc | cccc|}
    \cline{3-10}
       \multicolumn{2}{c|}{}& \multicolumn{4}{c|}{Retweeted user type} & \multicolumn{4}{c|}{Quoted user type}\\
    \cline{1-10}
        User type & Tot \% & Suspended &   Bots &   Non-bots &  Verified &   Suspended &   Bots &  Non-bots &   Verified \\
    \hline
        Suspended & 8.58\% & 7.55\%  & 1.18\%  &   48.35\% &    42.92\% & 7.31\% & 0.82\%   &   43.54\% &    48.33\% \\
        Bots & 5.6\% & 3.86\%  & 3.69\%  &   43.85\% &    48.6\% & 2.18\%  & 6.4\%  &   34.17\% &    57.25\% \\
        Non-bots & 84.51\% & 3.45\%  & 0.72\% &   54.11\% &    41.72\% & 3.02\% & 0.56\% &   47.19\%  &    49.23\% \\
        Verified & 1.32\% & 0.82\% & 0.47\% &   34.43\%  &    64.28\% & 1.25\% & 0.42\% &   26.06\% &    72.27\% \\
    \hline
    \end{tabular}
    \caption{\textbf{Percentage of retweets and quote tweets by users, divided by type.} Bots and suspended users are not very supportive of each other and their activity focuses on retweeting genuine accounts, thus generating noise and fostering discussions that are already present.}
    \label{tab:user_type_retweets}
\end{table}

\subsection{Core-periphery structure and the bots' position}
We analysed the position of bots and suspended users in the networks of retweets. We obtained each network of retweets by linking two users if one of them has retweeted the other at least once, without considering the direction of the retweet nor their number. In this sense, the resulting network will be undirected and binary. By not considering the number of retweets, we avoid noise from accounts whose activity is much higher with respect to the others.

\begin{figure}[htb!]
\centering
\resizebox{0.8\textwidth}{!}{%
 \includegraphics{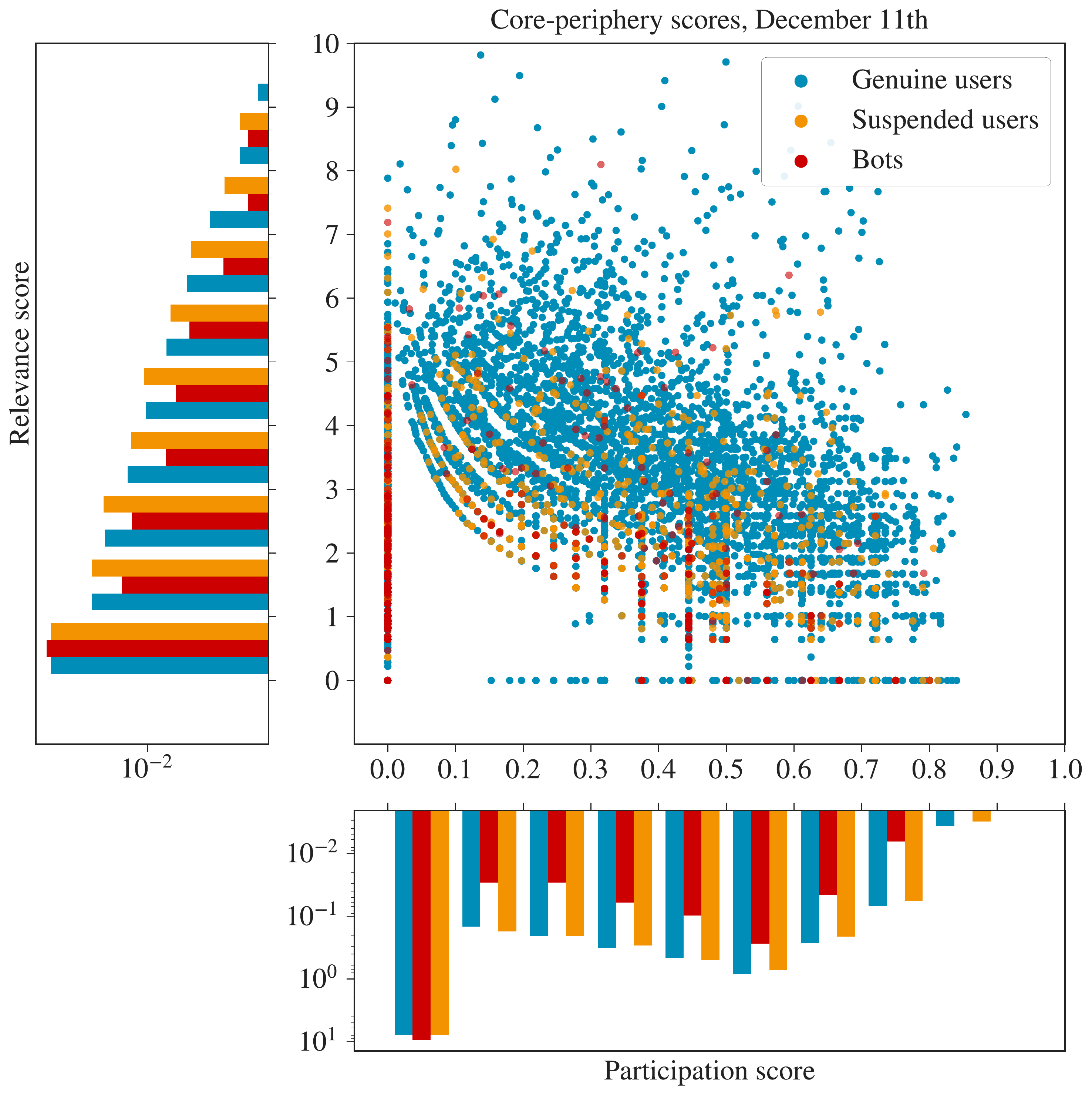}
}
\caption{\textbf{A phase diagram of the core-periphery scores with different colors for bots, suspended users and humans, for the network of retweets among users the day before the election.} The histograms on the bottom and left of the phase diagram show the marginal distributions. While automated accounts are concentrated on lower values of presence and participation scores, suspended and genuine users have a flatter distribution, even if a peak on the lower values is still present. Bots are more inclined to retweet accounts in their community (participation score) and moreover focus their activity on few users (relevance score).}
\label{fig:core_periphery}
\end{figure}

Our target is to understand whether the retweet activity of the bots is strategically determined to fuel echo chambers or to target undecided users, by analysing the position of the bots in the network. We partially follow the approach used in \cite{guimera2005} to measure the position of the nodes in the networks, and specifically whether the users lie in the core or periphery of their community. We employ a \emph{participation score} (Eq. \ref{participation_score}) to measure how much a user restricts its retweet activity on only one community, and a \emph{relevance score} (Eq. \ref{relevance_score}) to measure how relevant a user is inside its community, see the Methods section for more details. The results are shown in Fig. \ref{fig:core_periphery}: the bots seem to have a clear preference for targeting a single community with respect to the other users. In the figure, we can see the distributions for the 11th of December. Indeed, the distributions of the \emph{participation score} are very different when distinguishing between  genuine users and bots, with a seemingly higher preference of bots for retweeting only in their own community, thus having a lower participation. The \emph{relevance score} reveals instead a comparatively low centrality of the bots, meaning that the bots are usually on the periphery of one community, but focusing only on that single community with a low number of retweets, confirming the findings of \cite{gonzalez2021bots}. The low scores of bots in both participation and relevance, however, can also be dependent on the degree distribution as well, since they very often tend to limit their activity to a single or a few retweet.

\begin{figure}[ht!]
\centering
\resizebox{\textwidth}{!}{%
 \includegraphics{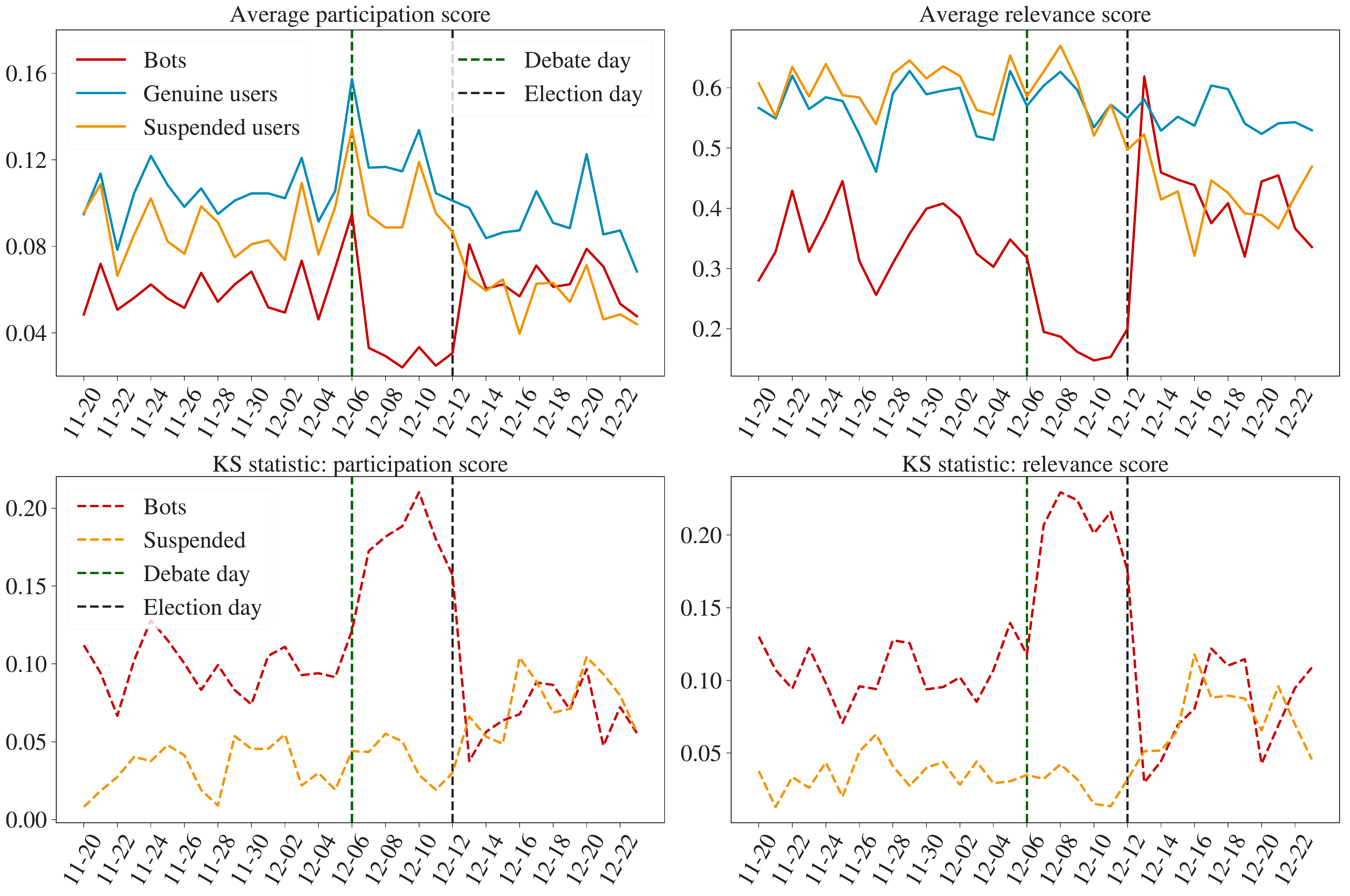}
}
\caption{\textbf{Average scores of participation and presence (top) by category and the KS tests' statistics comparing the distributions (bottom).} It is striking the change of behaviour of bots after the election day in both the presence and participation score: in fact, the average values of the scores are almost always higher for bots than for suspended users after the 12th of December. At the same time, the relevance score of suspended accounts is always higher than average users before the election day, dropping to much lower levels after the elections. The p-values of the KS tests are close to 0 during the respective peaks of the KS statistics, and almost always very small in the case of the bots (the maximum bots' p-value is $2\cdot 10^{-2}$).}
\label{fig:average_participation_presence_core_periphery}
\end{figure}

The difference between bots and users is well visible for the whole period when computing the average of the scores, as shown in Fig. \ref{fig:average_participation_presence_core_periphery}. The bots tend to limit their retweet activity to only one community, staying in the periphery of it. Interestingly, the signal intensifies during the week before the election, with a wider gap between bots and users. It is also worth noting that in this case, the suspended users present much higher scores than the bots and their average scores stay in line with genuine users until the date of the election, when the scores of suspended users become much closer to the bots' scores.

To quantify the difference between bots and users in the distribution of the core-periphery scores, we applied a two-sample Kolmogorov-Smirnov test. The results of the tests and their evolution over time can be seen in Fig. \ref{fig:average_participation_presence_core_periphery}, and confirm the significance of the difference for almost all days, and its increase  after the debate. 
For suspended users, a similar difference with respect to genuine users is found. After the increase in the percentage of suspended users, that is at the date of the election, the KS-tests of both measures yield a clear separation of the distributions of suspended and genuine users.

These results on core-periphery positions of the bots can also be explained by considering that the retweet activity of all users is not only focused on users of their same kind, as shown in Table \ref{tab:user_type_retweets}. Indeed, bots and non-bots  quote suspended users for a total of about 3-3.5\% of their activity, while suspended users retweet and quote other suspended users for more than 7\% of their total activity. A similar scenario appears for bots, whose retweet and quote activity focuses on bots much more (3.69\% and 6.4\%) than the activity of suspended and genuine users (around 1\%). Thus, the automated accounts do not do enough to push a consistent part of them to the core of the retweet networks, leaving them more to the periphery than the other users.

\subsection{Discursive communities}\label{sec:discursive_comms}
We analyzed the bipartite network of respectively verified and unverified users, obtained by linking a couple of users if one has retweeted the other at least once. Understanding the structure of such network is useful to characterize the behaviour of accounts with respect to pivotal topics and their interactions with different political sides. With such a network, we analyze the discursive communities that were introduced in the Methods section.

\begin{figure}[ht!]
\centering
\resizebox{0.7\textwidth}{!}{%
 \includegraphics{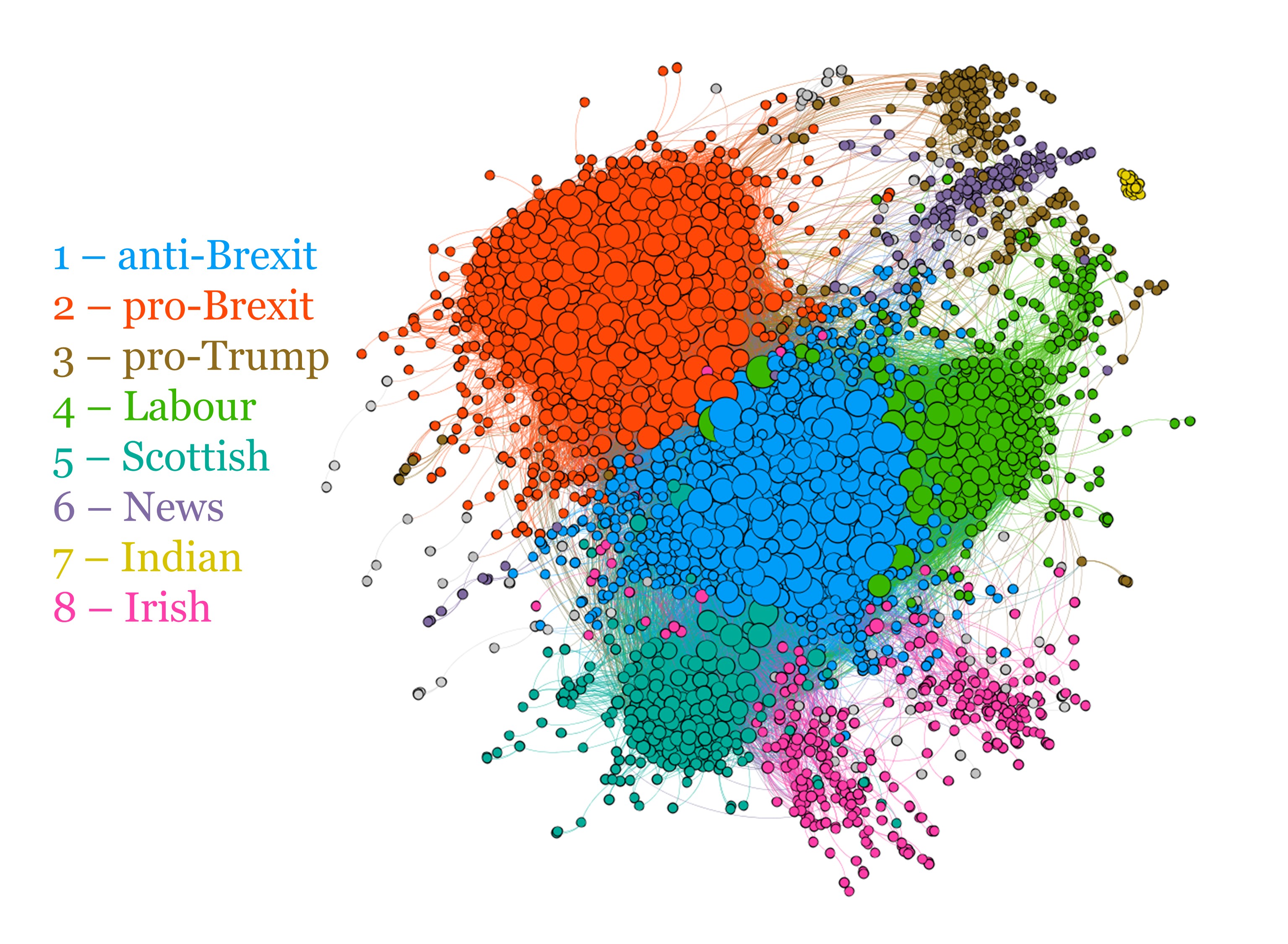}
}
\caption{\textbf{Projected network of the verified users.} The eight biggest communities have been highlighted and their composition is further explained in Table \ref{tab:discursive_comms}.}
\label{fig:discursive_comms}
\end{figure}

We first project the network on the layer of verified users by using the statistical projection described in Methods. We thus obtain a monopartite network of verified users (Fig. \ref{fig:discursive_comms}), that tells us which couples of verified accounts are retweeted by a significant number of common unverified users. With this network, we obtain labels for each verified user according to the community structure that we find by running the Louvain algorithm \cite{Blondel2008}. Then we are able to obtain the labels of unverified users by running a simple label propagation algorithm on the bipartite verified-unverified network. With the so found communities, we quantify the presence of bots and suspended users in each of them. 

The results of this procedure can be read in Table \ref{tab:discursive_comms}. The communities have very straightforward interpretations, each containing verified users that can be associated with a stance on Brexit or with a country such as India or Ireland. The biggest community is the one supporting EU and against Brexit, with mixed users involved, including actors and famous people, but also politicians and parties such as \textit{LibDems}. Then a pro-Brexit group includes the Conservatives and the Brexit Party. The pro-Brexit community has some interesting features. It presents a high percentage of suspended users, 11.76\%, that also appear to be organized as their validation \% is the highest among all groups. Moreover, the Brexit community has the highest link density (i.e. the ratio of the number of links over the number of users in the community) among all communities, meaning that the participation of users in this community is the highest. In general, the highest link densities are found in the groups that are directly involved with Brexit, i.e. pro- and anti-Brexit and the Scottish community; the Labour community link density is then surprisingly low. Interestingly enough, there is also a Trump supportive community, separated by the other groups and contains a staggering 23\% of suspended users and a high coordination of the automated accounts shown by the 12\% of validated bots, highest ratio among all groups, and the 7\% of validated suspended users, second highest. The Labour party is remarkably forming a community by itself (i.e. separated from the no-Brexit one), that includes also some news accounts. It makes sense to find the Labour community far from the pro-Brexit side but on its own: it highlights the fact that the stance on Brexit of the Labour party was rather unique, since they were proposing a renegotiation of the Brexit deal and a new referendum. Many news accounts lie in a community of their own, linked to communities of both pro- and anti-Brexit stances, but it is not rare to find some news accounts inside other groups. Other smaller main communities are related to nationalities: the Scottish, Indian and Irish communities can be found.

\begin{table}[!ht]
\setlength\tabcolsep{2pt}
    \centering
    \resizebox{\textwidth}{!}{
    \begin{tabular}{|c|c|c|c|c|c|c|c|c|c|}
    \hline
        \# & Type & Size & Most retweeted verified users & \begin{tabular}{c} Bots \\ \%\end{tabular} & \begin{tabular}{c} Suspended\\ \% \end{tabular} & \begin{tabular}{c} Verified\\ \% \end{tabular} & \begin{tabular}{c} Validated\\ bots \% \end{tabular} & \begin{tabular}{c} Validated\\ suspended \% \end{tabular} & \begin{tabular}{c} Link \\ density\end{tabular}\\
    \hline
        1 & anti-Brexit & 479718 & \begin{tabular}{c}
        davidschneider, DavidLammy, mrjamesob, \\ Femi\_Sorry, JimMFelton, HackedOffHugh, \\ Channel4News, Keir\_Starmer, LibDems,
        \end{tabular} & 4.66\% & 4.71\% & 1.33\% & 3.16\% & 4.68\% & 7.9
        \\
    \hline
        2 & pro-Brexit & 166254 & \begin{tabular}{c}
        BorisJohnson, Nigel\_Farage, LeaveEUOfficial, \\ Conservatives, brexitparty\_uk, darrengrimes\_, \\ GoodwinMJ, KTHopkins, JuliaHB1
        \end{tabular} & 6.91\% & 11.76\% & 1.03\% & 2.91\% & 7.34\% & 12.72
        \\
    \hline
        3 & pro-Trump & 112766 & \begin{tabular}{c}
        realDonaldTrump, ScottPresler, DiamondandSilk, \\ RealCandaceO, ddale8, greggutfeld, \\ DineshDSouza, IngrahamAngle, mtracey
        \end{tabular} & 3.7\% & 23.09\% & 0.71\% & 12.14\% & 7\% & 2.64
        \\
    \hline
        4 & Labour & 106227 & \begin{tabular}{c}
        jeremycorbyn, OwenJones84, BBCPolitics, \\ UKLabour, PeterStefanovi2,  PeoplesMomentum,\\ yanisvaroufakis, bbcquestiontime \end{tabular} & 8.79\% & 6.78\% & 0.75\% & 0.71\% & 1.92\% & 1.63
        \\
    \hline
        5 & Scottish & 20258 & \begin{tabular}{c} theSNP, dannywallace, joannaccherry, \\NicolaSturgeon, IrvineWelsh,
        Feorlean, \\ PeteWishart, AngusMacNeilSNP
        \end{tabular}  & 5.35\% & 4.24\% & 1,6\% & 1.94\% & 5.47\% & 5.78
        \\
    \hline
        6 & News & 13297 & \begin{tabular}{c}
        Reuters, BBCBreaking, business\\ Brexit, TheEconomist, AJEnglish, \\ AFP, nytimes, FT, CNN \end{tabular} & 8.92\% & 7.97\% & 4.93\% & 3.63\% & 1.13\% & 1.64
        \\
    \hline
        7 & Indian & 8908 & \begin{tabular}{c}
        gauravcsawant, swapan55, Iyervval,\\
        abhijitmajumder, AdityaRajKaul, TarekFatah,\\
        TVMohandasPai, WIONews, republic, samirsaran \end{tabular} & 1.95\% & 13.07\% & 0.67\% & 6.32\% & 4.3\% & 1.27
        \\
    \hline
        8 & Irish & 3323 & \begin{tabular}{c}
        SJAMcBride, naomi\_long, sinnfeinireland,\\ SenatorMarkDaly, GerryAdamsSF, ClaireHanna,\\
        Mr\_JSheffield, cstross, glynmoody, rtenews \end{tabular} & 5.09\% & 3.49\% & 11.95\% & 3.55\% & 0\% & 1.41
        \\
    \hline
    \end{tabular}
    }
    \caption{\textbf{Composition of the main discursive communities of the verified-unverified network after the label propagation.}. In Fig. \ref{fig:discursive_comms} the projection on the verified users is shown. The second and third columns from the right show the percentages of bots and suspended users in the community that appear in the hashtags-users projections. The higher these percentages are, the more the automated accounts are coordinated.}
    \label{tab:discursive_comms}
\end{table}

\subsection{What do the bots say?}

\begin{figure}[ht!]
\centering
\resizebox{\textwidth}{!}{%
 \includegraphics{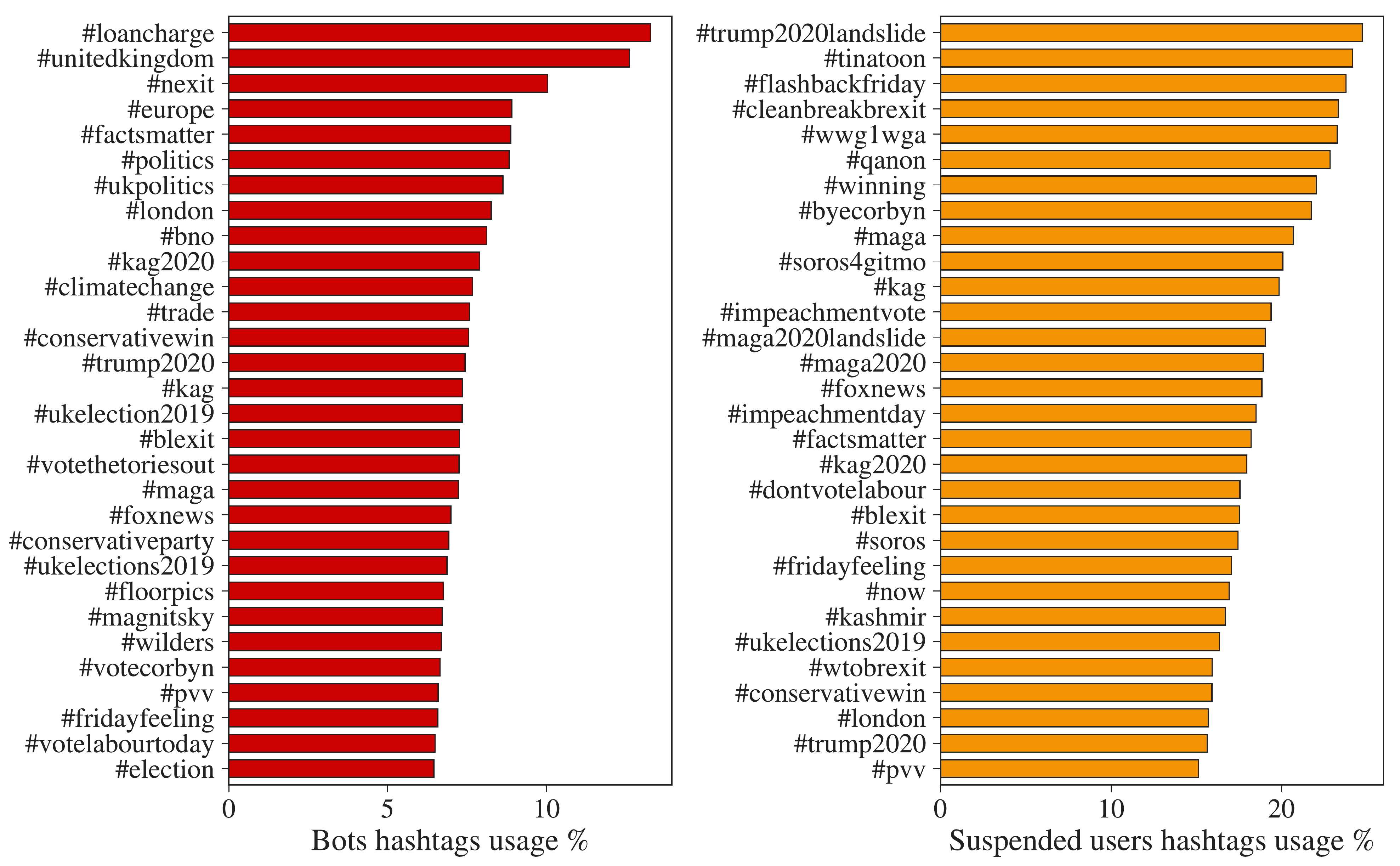}
}
\caption{\textbf{Most retweeted hashtags by bots and suspended users in percentage among all users' usage.} }
\label{fig:hashtags_histogram_retweet}
\end{figure}

To understand what the bots are posting about, we observed the use of hashtags and URLs by users in our dataset.
In Fig. 
\ref{fig:hashtags_histogram_retweet} we show the most retweeted hashtags (in percentage) by bots and suspended users.  
Among the most retweeted hashtags by suspended users, interestingly we can find topics that are external to Brexit, for example Trump 2020 campaign-related tags, such as \#Trump2020, \#MAGA,  \#QAnon, suggesting an external connection of the Brexit debate to Trump's campaign. Note that the use of popular hashtags such as \#fridayfeeling in combination with political topics was observed also in \cite{chowdhury2020twitter}. In the activity of bots, we still find Trump-related hashtags, even though we can see more focus on the actual theme of the election and hashtags related to all parties, such as \#votecorbyn, \#conservativewin and \#ukpolitics. It is also worth to mention the presence of \#nexit (the word Nexit is the equivalent of Brexit for the Netherlands) and there are similar hashtags related to other countries.

\begin{figure}[ht!]
\centering
\resizebox{\textwidth}{!}{%
 \includegraphics{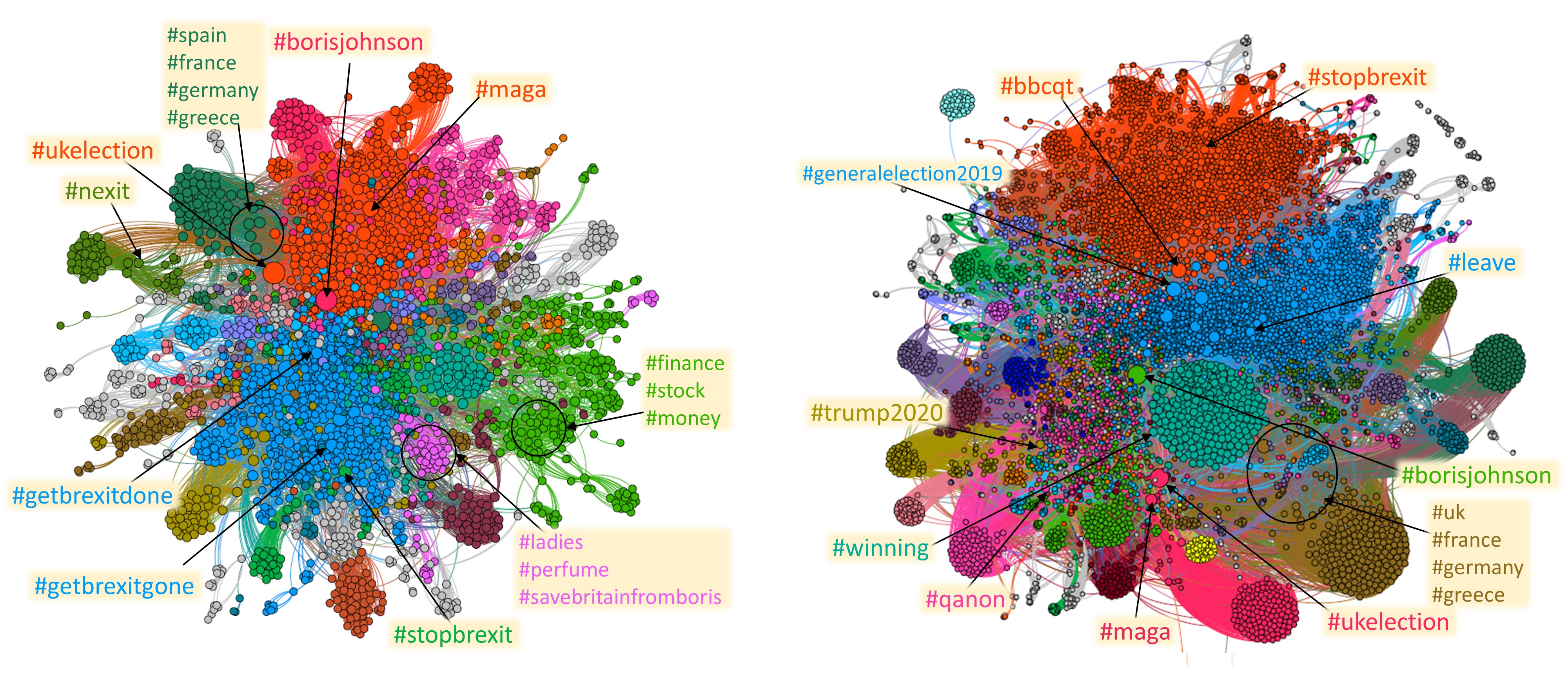}
}
\caption{\textbf{The backbone of the network of hashtags and bots (left) or suspended users (right) linked by retweets for the whole period of our dataset.} Both networks show a modular structure probably due to the coordination of the automated users: the depicted partitions have a modularity of 0.73 (left) and 0.78 (right). In the bots' network the Brexit discussion appears together in the blue community, while for the suspended users two separate groups are pro-Euro (orange) and pro-Brexit (blue). In both cases, Trump-related hashtags are very common.}
\label{fig:bots_hashtags_networks}
\end{figure}

To improve our understanding of the bots organization, we built the bipartite networks connecting, respectively, bots and hashtags, and suspended users and hashtags. A connection is present between an account and a hashtag if the latter was retweeted at least once by the user. A first inspection suggests a separation of the activity of bots or suspended users in groups. In order to strengthen the signal, we extracted a backbone of the networks by projecting the bipartite network on its layers and re-linking the nodes of the two monopartite projections with the original links (see the Methods section for details). The two backbone networks can be seen in Fig. \ref{fig:bots_hashtags_networks}, with  bots-hashtags on the left, and suspended-hashtags on the right.
Our methodology reveals the dense cliques of hashtags that have common retweeters among bots and suspended users. We then search for communities in the network by running the Louvain algorithm for modularity optimization.

Among the mixed suspended users-hashtags communities highlighted in the right-side network of Fig. \ref{fig:bots_hashtags_networks}, there are some that are naturally linked together, and some that are most likely corresponding to an unusual activity of bots. For example, the orange and the blue communities, that are the largest ones, contain hashtags that refer to the elections and each one leans towards one direction (conservative/Brexit for the blue one, labour/pro-EU for the orange one). Meanwhile, some smaller but denser communities hashtags related to Qanon and the Trump campaign (pink, golden and purple, bottom left) or contain links to other EU countries such as France, Germany and Greece.
For the network of bots, on the left of Fig. \ref{fig:bots_hashtags_networks}, the situation is similar but this time the discussion on Brexit is mixed in one community (blue), while the other communities contain either groups similar to those found for the suspended users, or other generic topics, maybe injecting noise or advertising products.

\begin{figure}
\centering
\resizebox{\textwidth}{!}{%
 \includegraphics{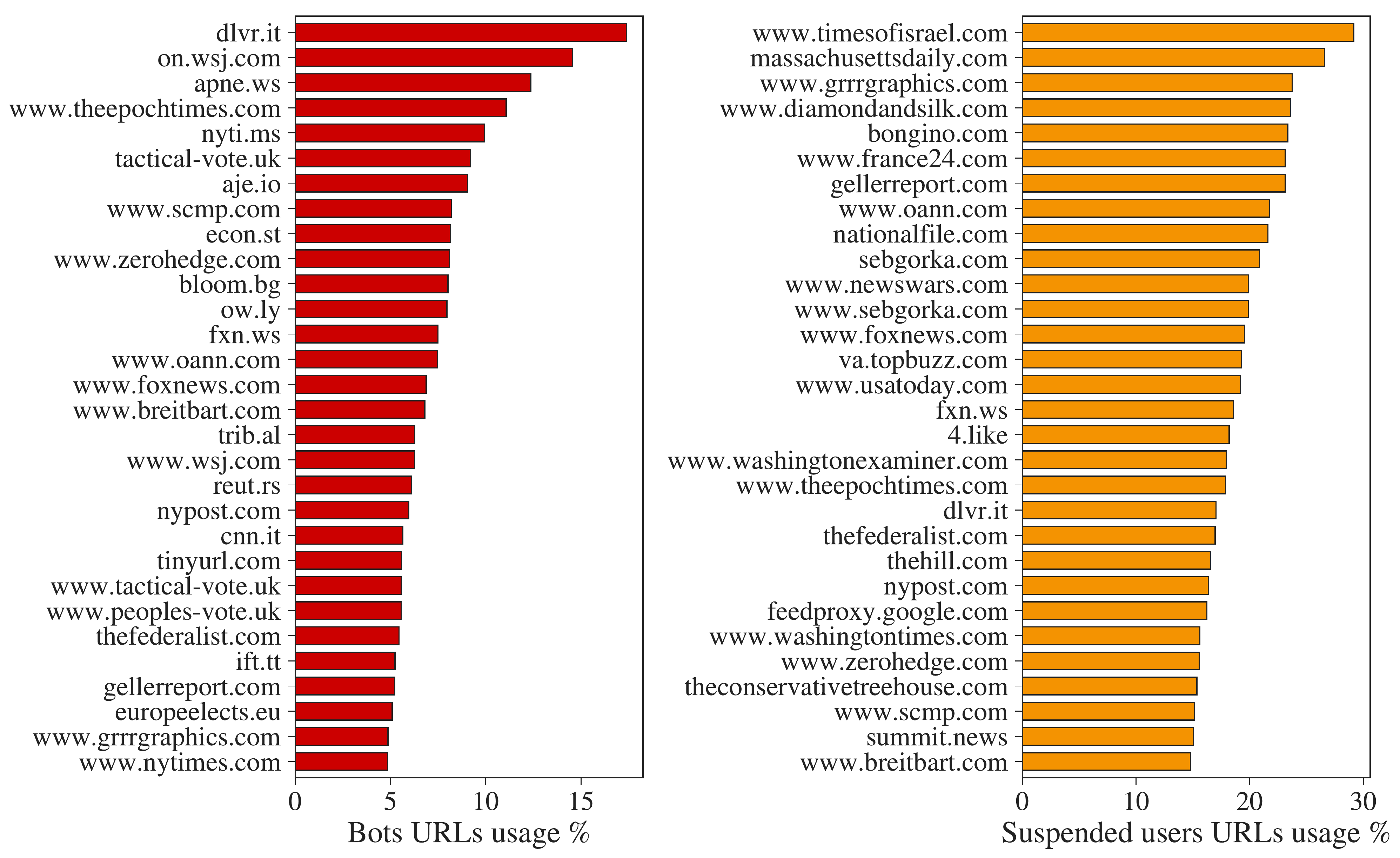}
}
\caption{\textbf{Most retweeted URLs by bots and suspended users in percentage among all users' usage.}}
\label{fig:urls_histogram_retweet}
\end{figure}

\begin{figure}
\centering
\resizebox{\textwidth}{!}{%
 \includegraphics{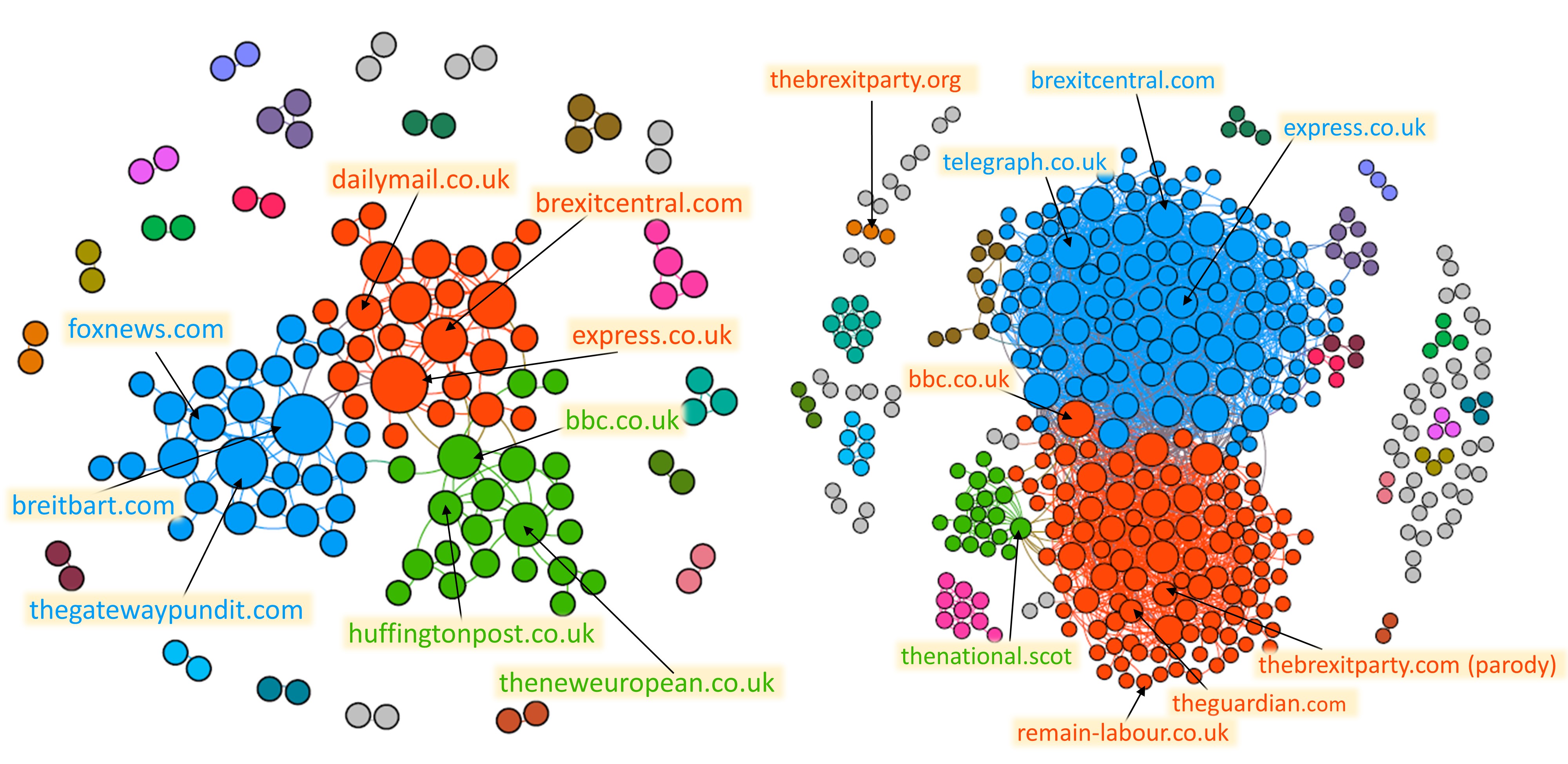}
}
\caption{\textbf{The projected networks of URLs used by bots (left) and by suspended users (right) for the whole period we consider.} The suspended users are more organized in their patterns of retweet of URLs. Interestingly, Trump-related websites appear only in the bots' projection.}
\label{fig:bots_urls_networks}
\end{figure}




Then we analysed the use of URLs by bots and suspended users. In Fig. 
\ref{fig:urls_histogram_retweet} we present the URLs most used by bots and suspended users compared to genuine users. The retweet activity of both categories, but particularly suspended users, spams many websites that are close to Trump's campaign, such as \emph{diamondandsilk.com}, \emph{gellerreport.com}, \emph{breitbart}, \emph{grrrgraphics}.

We also built the bots-URLs retweet network (Fig. \ref{fig:bots_urls_networks}) similarly to what we did for the hashtags, and in the same spirit we projected the network on the layer of the URLs via a statistical projection. 
Interestingly, on the projection from the URL-suspended users network we can find a separation of the URLs in the pro-EU and pro-Brexit links, and that is reflected also in newspapers URLs, with the most conservative ones leaning towards the blue community while the most pro-EU lie in the orange community. The green community on the right-side network in Fig. \ref{fig:bots_urls_networks} is mostly made of Scottish websites. For the bots-URLs projection (left-side network of the same figure), although the projection is less populated, the story is similar, with a separation in pro-Euro (green) and pro-Brexit (blue) groups, but this time there is a community of Trump supportive websites of comparable size, including URLs and disinformation websites such as \emph{diamondandsilk.com}, \emph{gellerreport.com}, \emph{breitbart.com} and such. This is analogous to what happened in the networks of bots/suspended users and hashtags. In this case, the projection on the users' layer did not validate links between bots or suspended users since the use of URLs by users is way less diversified. 


\section{Discussion and conclusion}

Our analysis confirms that bots played a central role in the Brexit Twitter discussion. We have observed that the percentage of bots went from less than 2\% to more than 11\% in one week, that is crucially the one just before the elections. This radical increase is observed right after the TV debate between the leaders of the main political parties, Boris Johnson and Jeremy Corbyn, which happened six days before the election. Such findings suggest that the injection of bots in the Brexit discussion was not a coincidence, but had the purpose of amplifying the importance of Brexit as a topic during the general  election. Unfortunately, due to the limited time-window of our data, we cannot say if these bots are new to the whole political discussion or if they were simply shifting their attention on Brexit.
Furthermore, we have observed differences in the temporal patterns of bots and suspended users. The presence of the suspended users is not increasing at the same time of bots, suggesting that one wave of bots went under the radar in Twitter's bans, and that different types of automated accounts populate the Twitter landscape.

We were able to highlight the differences in their activity by analysing the different positions of bots and humans in the network structure. We started considering the core-periphery structure of the network, using the approach of \cite{guimera2005}: we analysed the position of the bots in the communities found in the networks of retweets, finding that automated accounts tend to be embedded in one single community and to stay in the periphery of it, confirming previous findings~\cite{stella2018bots}. This is in part due to the large portion of bots having low degree, but also to the small portion of bots retweeting bots, thus not being able to push some of the bots to the core of communities. We also showed that the statistical differences between bots and humans increased with the mass insertion of bots.

By considering the retweet patterns of unverified users towards verified users, we were able to highlight  groups of users tweeting about similar topics, and we analyzed the participation of bots and suspended users in such groups. We found that the activity of bots is not limited to specific discussions, but instead they are present everywhere, although in different proportions. For suspended users instead, we found that a large part of their activity is focused on the Trump campaign. 
Moreover, the analysis of the hashtags and URLs usage by the bots has revealed that while there was a big activity by automated accounts, the streams of coordinated bots  do not look to be only specifically linked to the Brexit discussion, but also on different topics that could benefit from the Brexit debate, mainly populist narratives. The two main topics that we uncovered are the Trump 2020 campaign, and perspectives of a further division in the EU.

Summarising, in this paper, we have found a significant presence of bots in the Brexit discussion during the UK elections of 2019, following the findings of Bastos and Mercea \cite{bastos2017} and Howard and Kollanyi \cite{howard2016bots} that also found a large presence of bots during the Brexit referendum. The activity of bots also presents a steep, apparently coordinated, increase, and this result is robust to the choice of the threshold used in the Botometer score. This increase comes in a crucial time, after a widely popular TV debate between political candidates one week before the elections and the elections themselves. We further analysed the behaviour of the automated accounts by analysing several network features, such as the core-periphery positions of users and the similarities of bots in the retweet patterns via the analysis of different kind of networks. We found that the automated accounts are spread across all the discussions. Furthermore, the bots behave significantly differently with respect to the genuine users, and we found groups of bots retweeting hashtags not always perfectly centred on the Brexit discussion but relative to other populist narratives.
Due to the nature of our dataset, some questions  remain unanswered. For instance, we do not know if the spike in bots' activity is due to political bots shifting their attention on Brexit, as that would require a more general stream of tweets. Our findings suggest that at times the behaviour of bots is common across political discussions, and generating noise that is not particularly polarised can be used as a tool by multiple sides. However, further similar analyses on the specialisation of bots on a certain subtopic of a general discussion are needed to investigate these questions.

\bibliographystyle{unsrt}

\appendix
\section{Choice of the CAP threshold in Botometer}\label{app:cap_discussion}
\begin{figure}[ht!]
\centering
\resizebox{0.8\textwidth}{!}{%
 \includegraphics{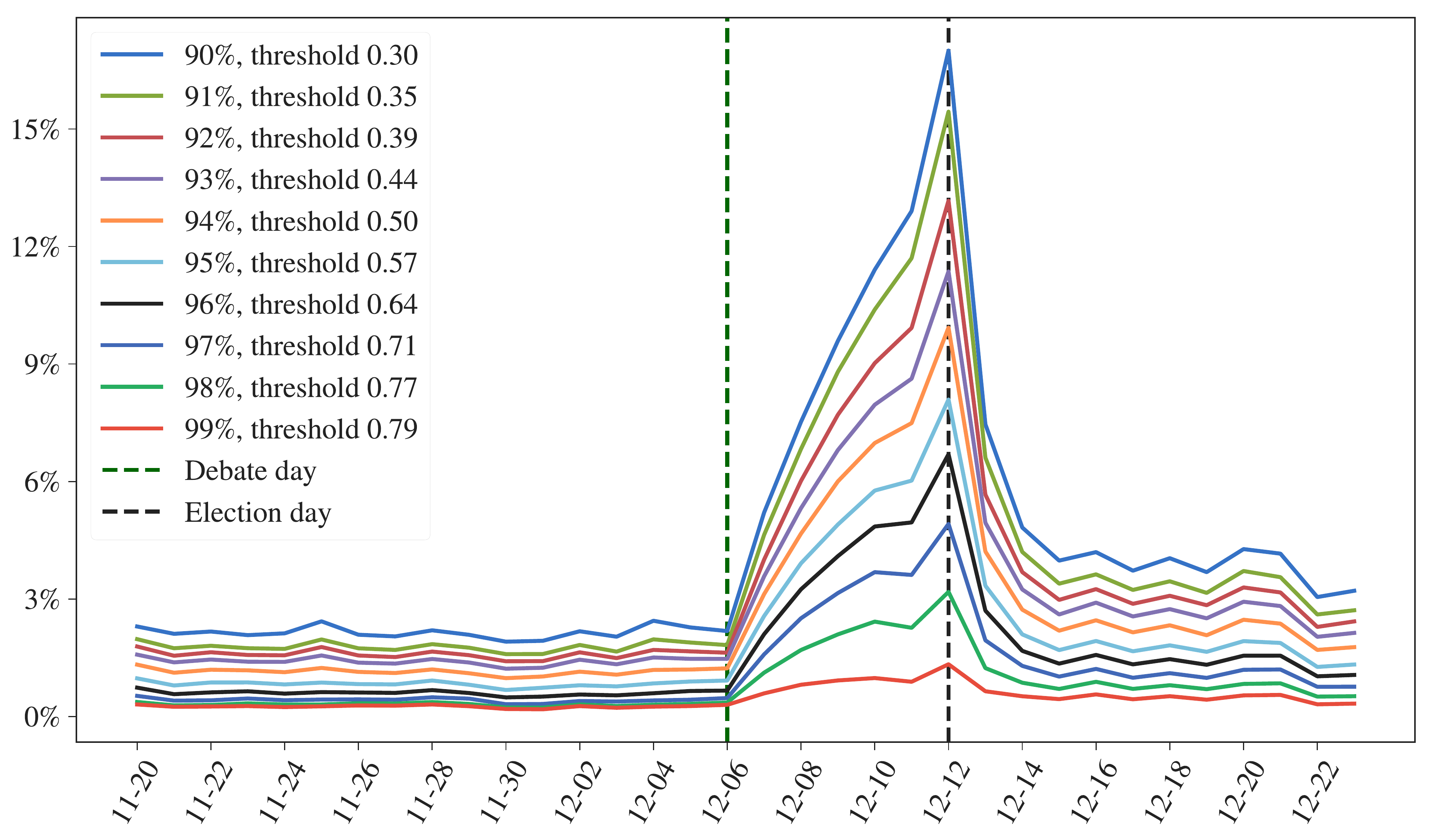}
}
\caption{\textbf{Percentage of bots per day cutting setting different thresholds, accounting for the top 1-10 percentiles}. The relative increase in bots' presence is independent of the chosen threshold.}
\label{fig:bots_percentiles}
\end{figure}

It is remarkable that the pattern observed in Fig.~\ref{fig:bots_percentage} is independent of the threshold chosen to label users as bots: in Fig. \ref{fig:bots_percentiles} we show that the relative percentage increase of bots happens with different thresholds, chosen statistically to label as bots the top 1-10 percentiles. In our case, we chose a threshold that was already used and labels as bots about 7 percent of all users in the dataset.

\section{Behaviour of incoming bots}

\begin{figure}
\centering
\resizebox{0.8\textwidth}{!}{%
 \includegraphics{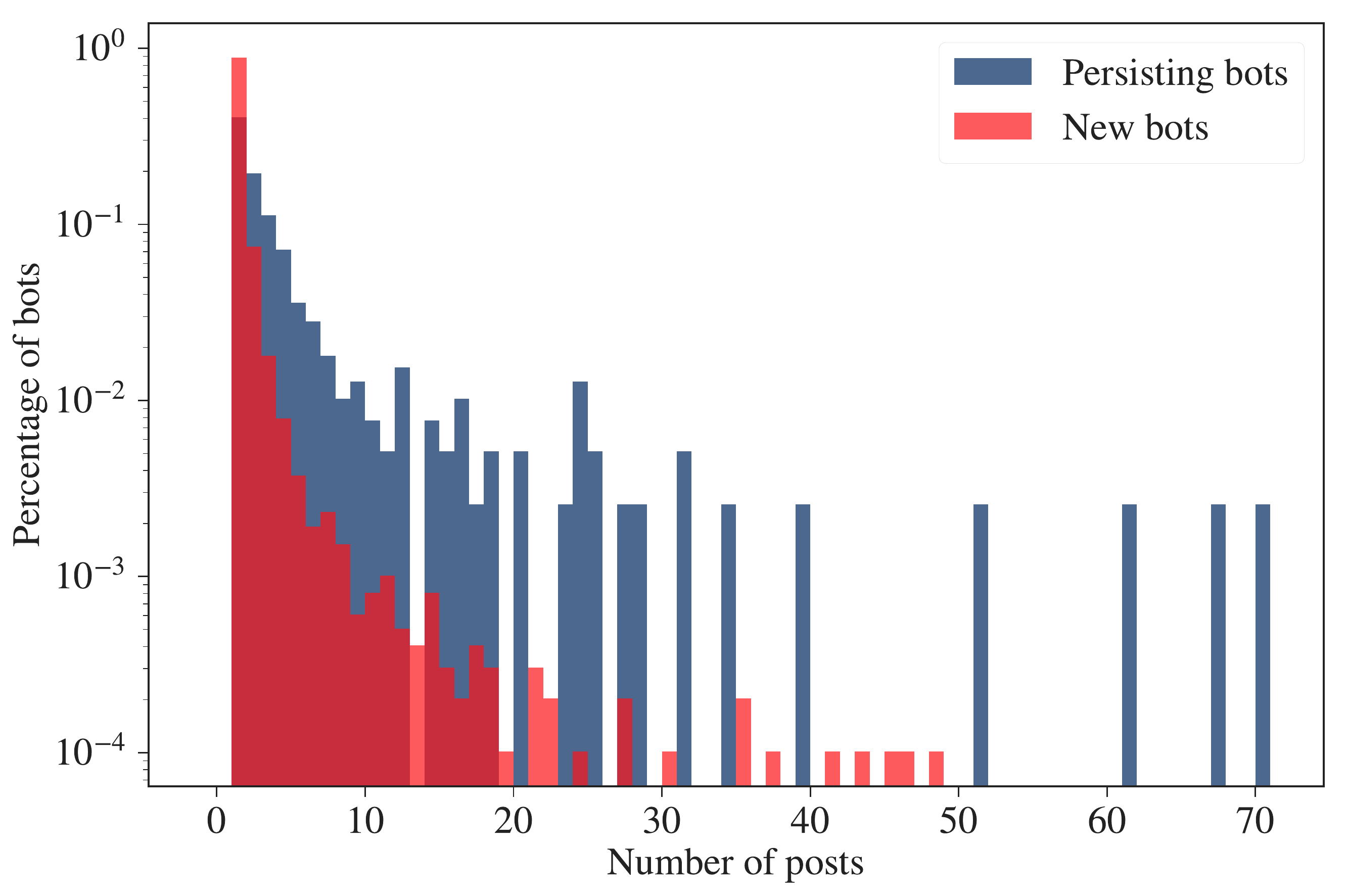}
}
\caption{\textbf{Histogram of the number of posts made by bots.} We took bots from the 11th of December, and the two histograms separate the activity of those that were already present in the discussion the 6th of December (persisting bots) from the new entries (new bots). Almost 40\% of those of the 6th of October are still in the discussion, and their average activity is almost stable from the two dates. The old users' activity is much higher, so the effect of the less activity is only caused by the new bots' behaviour.}
\label{fig:bots_activity_histogram}
\end{figure}

It could be argued that the change of behaviour of the bots is not caused by the incoming automated accounts but rather by a change of behaviour of the ones that were already present in the dataset. Fig. \ref{fig:bots_activity_histogram} suggests that this should not be the case. To explore this hypothesis, we separated the bots present on the 11th of December between the ones that were also present before the 6th of December, and the ones that were not. New bots are clearly less active, usually limiting their activity to a single retweet, while the ones that were already in the discussion show a more variegated activity.

\section{Automated retweeters of verified users}

\begin{figure}
\centering
\resizebox{\textwidth}{!}{%
 \includegraphics{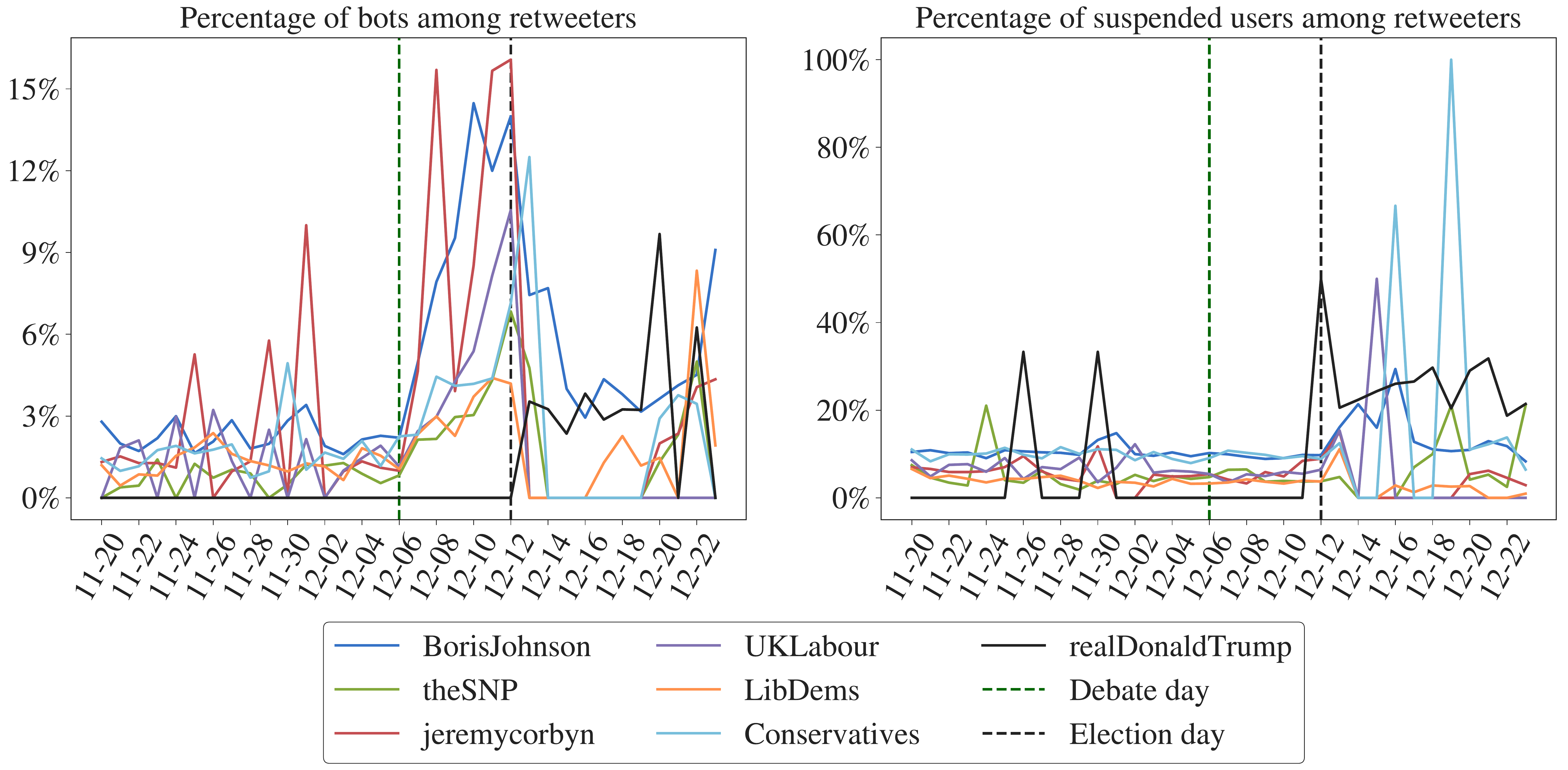}
}
\caption{\textbf{Percentage of bots and suspended users among retweeters for some main verified accounts.} The percentages of suspended accounts increase after the election day, while in the case of bots the increase is observed after the day of the election.}
\label{fig:retweeters_bots}
\end{figure}

We also analysed what is the impact of bots and suspended users on the retweeters of some of the most popular verified accounts from various political parties. We found that the increase of bots is transversal across the users that we considered, while the suspended users' increment seems to be driven by mainly \emph{realDonaldTrump}'s retweeters, and partially \emph{BorisJohnson}'s retweeters (Fig. \ref{fig:retweeters_bots}).

\end{document}